\RequirePackage{chemformula}

\documentclass[pdflatex,default]{sn-jnl}

\usepackage{siunitx}

\jyear{2022}

\begin{document}

\title{Microjoule-level mid-infrared femtosecond pulse generation in hollow-core fibres}

\author[1]{\fnm{Ang} \sur{Deng}}
\author[1]{\fnm{Trivikramarao} \sur{Gavara}}
\author[1]{\fnm{Muhammad Rosdi Abu} \sur{Hassan}}
\author[2]{\fnm{Md Imran} \sur{Hasan}}
\author*[1]{\fnm{Wonkeun} \sur{Chang}}\email{wonkeun.chang@ntu.edu.sg}

\affil[1]{\orgdiv{School of Electrical and Electronic Engineering}, \orgname{Nanyang Technological University}, \orgaddress{\street{50 Nanyang Avenue}, \postcode{639798}, \country{Singapore}}}
\affil[2]{\orgdiv{Research School of Physics}, \orgname{The Australian National University}, \orgaddress{\city{Canberra}, \postcode{2601}, \state{ACT}, \country{Australia}}}

\abstract{We demonstrate a fibre-based approach that generates mid-infrared femtosecond pulses in the \SIrange[range-phrase=--,range-units=single]{3}{4}{\micro\meter} spectral region with microjoule-level single pulse energy. This is realised in a piece of gas-filled antiresonant hollow-core fibre that is pumped by a two-micron light source. A rapid variation of the dispersion near a structural resonance of the fibre creates a phase-matching point in the mid-infrared, which mediates the frequency-down conversion. We generate femtosecond pulses centred at \SI{3.16}{\micro\meter} wavelength with the pulse energy of more than \SI{1}{\micro\joule}, achieving the conversion efficiency as high as \SI{9.4}{\percent}. The wavelength of the radiation is determined solely by the dielectric wall thickness of the cladding elements, while the yield is subject to other experimental parameters. This, combined with high power-handling capability of hollow-core fibres, makes it possible to power scale the mid-infrared output by either increasing the pulse energy or repetition rate of the pump. The technique presents a new pathway to build an all-fibre-based mid-infrared supercontinuum source, which promises to be a powerful new tool for ultrahigh sensitivity molecular spectroscopy.}

\maketitle

\section{Introduction}

High-power ultrafast light sources have become an indispensable tool in science and technology. Most powerful laser systems generating femtosecond pulses operate around one-micron wavelength with available average output power in the kilowatts level \cite{Mueller:2020:OL}. The two-micron region is also gaining a lot of attention lately, emerging as a new wavelength for high-throughput ultrafast lasers \cite{Gaida:2018:OL}. Towards the longer wavelength, however, the available power drops dramatically. The lack of powerful ultrafast sources in mid-infrared (mid-IR) stands in stark contrast to their many applications. Highly concentrated mid-IR photons are attractive for triggering extreme nonlinear effects such as high-harmonic generations \cite{Popmintchev:2012:Science}. Moreover, they can efficiently drive coherent mid-IR supercontinuum generation \cite{Petersen:2014:NP}, offering a unique opportunity in molecular spectroscopy \cite{Ebrahim-Zadeh:2008:Book}. With a powerful mid-IR supercontinuum source, the ``fingerprints'' of various molecules can be identified in a single measurement at an unprecedented sensitivity.

A common solution to generate femtosecond mid-IR pulses rely on optical parametric processes in nonlinear crystals \cite{Seidel:2018:SciAdv,Elu:2017:Optica}. While the technique is largely successful, delivering multi-watt-level average output power, it has a major drawback of being prone to vibrations and other environmental variations. Fibre-based systems have the advantage in this regard, but silica---commonly used material for optical fibres---is not transparent beyond \SI{2.4}{\micro\meter}. A recent progress made in modelocked lasers incorporating erbium-, holmium-, or dysprosium-doped fluoride fibre yielding nanojoule pulses \cite{Huang:2020:Optica,Bawden:2021:OL,Antipov:2016:Optica,Wang:2019:OL} marks a breakthrough in the ultrafast mid-IR light source development.

Another platform that is attracting growing interest in mid-IR photonics is hollow-core fibres \cite{Wheeler:2014:OL,Yu:2019:APLPhton}. In a hollow-core fibre, the light is guided in its central hollow region, bypassing the limitations imposed by intrinsic properties of the waveguide material. It enables low-loss mid-IR transmission even in silica-based optical fibres. An interesting prospect is then to use hollow-core fibre as a container for matter particles to induce lasing at or frequency conversion to mid-IR wavelengths \cite{Xu:2017:OL,Aghbolagh:2019:OL,Zhou:2022:LSA,Adamu:2019:SR,Deng:2021:CLEO}. Many demonstrations have been made in the last decade on the use of gas-filled kagom\'e-lattice, antiresonant, or simple capillary hollow-core fibres to achieve phase-matched conversion of near-infrared pump to, mostly, the ultraviolet region \cite{Joly:2011:PRL,Mak:2013:OE,Koettig:2017:Optica,Travers:2019:NP,Xiong:2021:PR}. This is due to general landscape of the dispersion in these fibres, which permits soliton-dispersive wave phase matching only in the blue-side. It is reported that phase matching to mid-IR is possible in gas-filled hollow-core fibre by exploiting the photoionisation effect, albeit with rather poor efficiency \cite{Koettig:2017:NC}.

One promising approach to generate mid-IR radiation is to utilise so-called the band-edge effect in antiresonant hollow-core fibres (AR-HCFs) \cite{Tani:2018:PR}. It exploits the effect of the structural resonances in AR-HCF, which dominates and overrules the dispersion profile around them. The concept has been tested in a recent experiment, successfully frequency-downshifting the pump at \SI{800}{\nano\meter} to \SI{1.45}{\micro\meter}, i.e., to a wavelength determined by the fibre geometry \cite{Gavara:2020:OL}. Here we build on this idea to realise a fibre-based ultrafast mid-IR light source. Namely, we produce microjoule-level femtosecond pulses centred at \SI{3.16}{\micro\meter} that emerge from an argon-filled AR-HCF pumped at \SI{2}{\micro\meter} with the conversion efficiency as high as \SI{9.4}{\percent}. We also investigate the role of fibre geometry in the process by employing another fibre of different dimensions, demonstrating shifting of the mid-IR radiation to \SI{3.61}{\micro\meter}.

\section{Results}

\begin{figure}[h]
  \centering
  \includegraphics[scale=0.66]{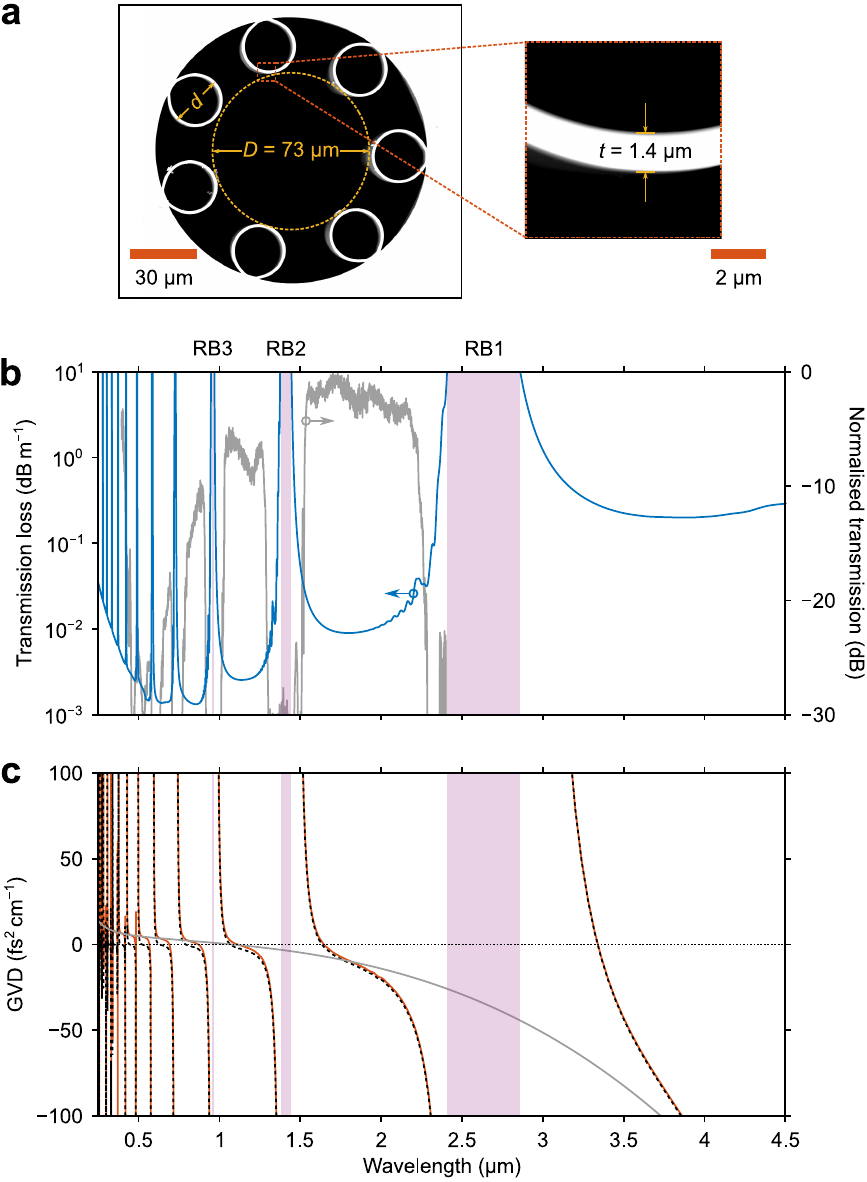}
  \caption{\textbf{AR-HCF characterisation.} \textbf{a.} A cross-sectional image of the AR-HCF taken with a scanning-electron microscope. The right-hand side is the magnified image of a cladding element. The fibre has core diameter $D=\SI{73}{\micro\meter}$, average outer cladding element diameter $d=\SI{24}{\micro\meter}$, and dielectric wall thickness $t=\SI{1.4}{\micro\meter}$. \textbf{b.} Calculated transmission loss of an idealised AR-HCF of the same geometry (blue-solid line), and normalised transmission spectrum taken at the output of a \SI{2}{\meter}-long fibre (grey-solid line). \textbf{c.} Calculated GVD of the AR-HCF when evacuated (black-dotted line) and argon pressurised at \SI{16}{\bar} (red-solid line). GVD of a simple dielectric capillary of the same core size and argon pressurisation is also plotted for comparison (grey-solid line). RB1, RB2, and RB3, denote the first three $t$-induced resonant bands (magenta shades).}\label{fig1}
\end{figure}

The frequency conversion is staged in a gas-filled AR-HCF. Figure \ref{fig1}a presents a cross-section of the fibre used in the experiment. It consists of seven silica tubular cladding elements, which have average outer diameter $d=\SI{24}{\micro\meter}$ and wall thickness $t=\SI{1.4}{\micro\meter}$. The light is guided in the hollow core of diameter $D=\SI{73}{\micro\meter}$. Inhibition of the coupling between spatial modes in the hollow core and those in the cladding ensures that light launched in the core is confined and transmitted along the fibre with minimal leakage \cite{Debord:2017:Optica,Deng:2020:OE}. It turns out dielectric modes in the cladding wall interact the most strongly with the core modes, making the wall thickness $t$ one of the key geometrical parameters in the AR-HCF guidance. Namely, $t$ governs resonant bands where light in the core leaks out through coupling to the dielectric modes. This occurs at wavelengths given by \cite{Archambault:1993:JLT}:
\begin{equation}
  \lambda_{m}=\frac{2t}{m}\sqrt{n^{2}-1}\textrm{,}\label{eq1}
\end{equation}
where $n$ is refractive index of the dielectric material, i.e., silica, and $m=1$, $2$, $3$, ... is the resonance order. The hollow guidance in the core then happens in low-loss regions between the resonances.

Figure \ref{fig1}b shows calculated transmission loss of the fundamental core mode of an idealised AR-HCF that has the same cross-sectional dimensions as the one shown in Fig.~\ref{fig1}a. The presence of the resonant and antiresonant bands across the spectrum is evident. The normalised transmission spectrum taken at the output of a \SI{2}{\meter}-long fibre is also plotted in Fig.~\ref{fig1}b. It matches well with the calculated loss spectrum, as well as locations of the resonances given by Eq.~(\ref{eq1}). The first three loss bands appear in \numrange[range-phrase=--]{2.41}{2.86}, \numrange[range-phrase=--]{1.38}{1.44}, and \SIrange[range-phrase=--,range-units=single]{0.95}{0.97}{\micro\meter}, respectively. Hence, the \SI{2}{\micro\meter}-pump lies in the second transmission window of the fibre, and the first window covers mid-IR with the calculated loss of less than \SI{1}{\decibel\per\meter} in the \SIrange[range-phrase=--,range-units=single]{3}{5.2}{\micro\meter} range. The transmission loss of less than \SI{0.35}{\decibel\per\meter} is measured at \SI{2}{\micro\meter} using the cut-back method.

The resonances also largely influence group-velocity dispersion (GVD) in their vicinity as shown in Fig.~\ref{fig1}c. In the central part of each transmission band, GVD closely follows that of a simple dielectric capillary of the same core size and gas pressurisation \cite{Marcatili:1964:BSTJ}. However, it deviates and varies rapidly near the band edges. As we shall see, this dramatic GVD change in the band edges is the key to the phase-matched frequency conversion to mid-IR. Note that in the band edges, the $t$-induced resonances dominate the dispersion, while the gas filling makes only a minor impact. The system exhibits anomalous dispersion at \SI{2}{\micro\meter} when pressurised with argon at \SI{16}{\bar}, which permits the formation of optical solitons.

\begin{figure}[h]
  \centering
  \includegraphics[scale=0.66]{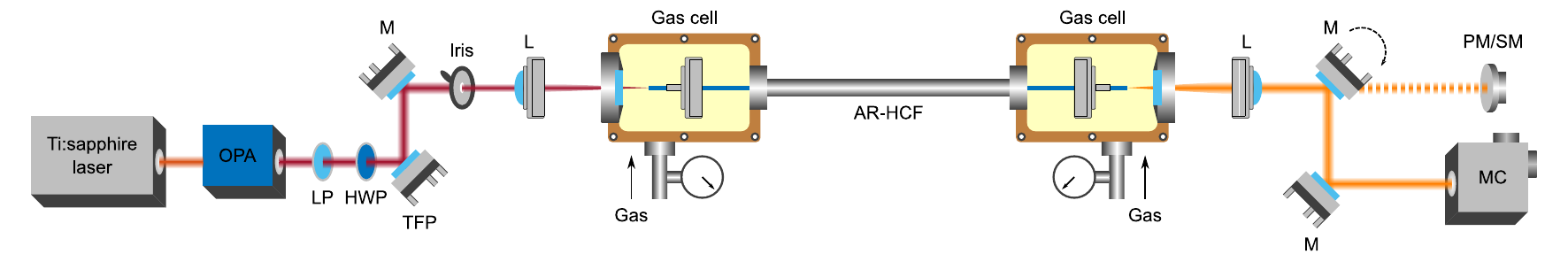}
  \caption{\textbf{Experimental setup.} A schematic of the experiment. OPA, optical parametric amplifier; LP, long-pass filter; HWP, half-wave plate; TFP, thin-film polariser; PM, power meter; SM, spectrometer; MC, monochromator; M, mirror; L, lens.}\label{fig2}
\end{figure}

The experimental setup is illustrated in Fig.~\ref{fig2}. A \SI{25}{\centi\meter}-long AR-HCF is pressurised with argon. One of the two fibre ends is pumped by \SI{2}{\micro\meter} idler beam from an optical parametric amplifier that is driven by a Ti:sapphire laser operating at \SI{1}{\kilo\hertz} repetition rate. The pump pulse has \SI{65}{\femto\second} full width at half maximum (FWHM) duration (See Supplementary Note 1). The beam exiting from the output end of the fibre is focused to diagnostics.

\begin{figure}[h]
  \centering
  \includegraphics[scale=0.66]{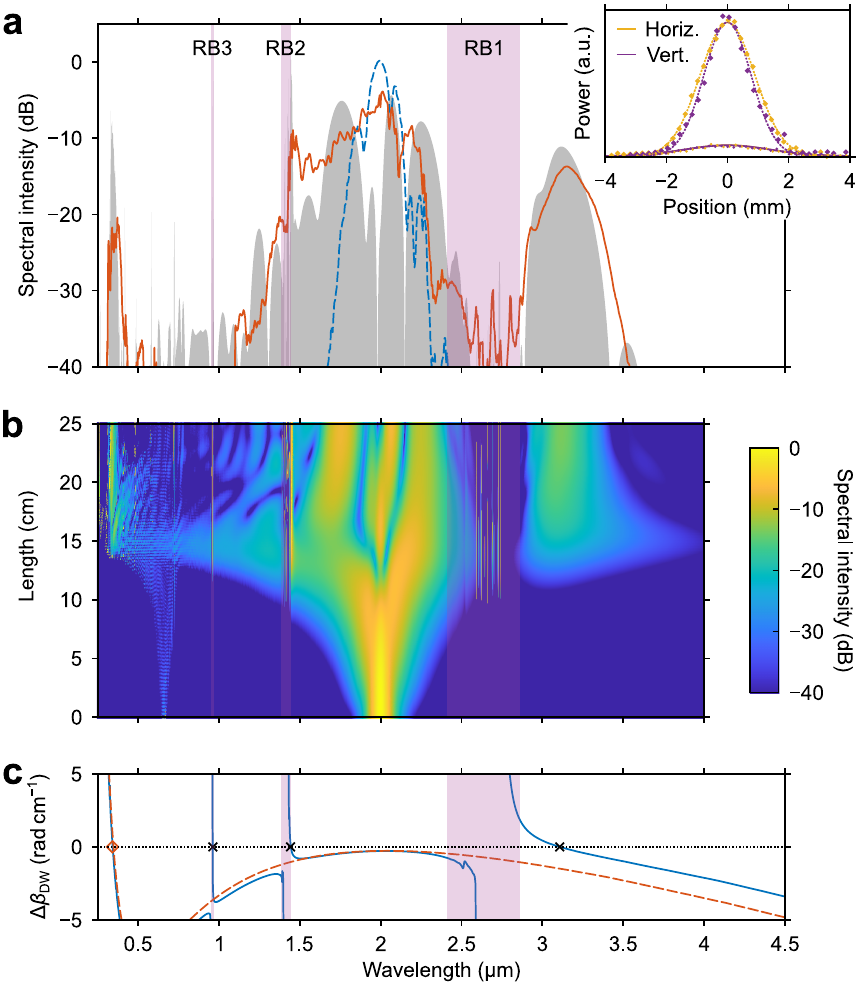}
  \caption{\textbf{Mid-IR generation.} \textbf{a.} Measured (red-solid line) and simulated (grey shade) spectra at the AR-HCF output. The fibre is pressurised with \SI{16}{\bar} argon and the pump pulse energy is \SI{12.5}{\micro\joule}. The pump spectrum is plotted together for reference (blue-dashed line). The inset is intensity profiles of the total (dotted lines) and mid-IR (solid lines) beams emerging from the fibre output characterised using the knife-edge method. \textbf{b.} Simulated spectral evolution along the fibre length. \textbf{c.} The corresponding soliton-dispersive wave phase-matching diagram (blue-solid line). The dephasing of a simple dielectric capillary of the same core size and argon pressurisation is included for comparison (red-dashed line). Phase-matching points that originate from the resonant bands (black crosses) and capillary model (red diamond) are marked in the diagram. Only the first three resonant bands, denoted RB1, RB2, and RB3 (magenta shades), are considered in the dephasing diagram and band-edge induced phase-matching points for clarity.}\label{fig3}
\end{figure}

Figure \ref{fig3}a presents an output spectrum when the fibre is pressurised with \SI{16}{\bar} argon and the energy of the pulse exiting the fibre is \SI{11.7}{\micro\joule}. The corresponding pump pulse energy is \SI{12.5}{\micro\joule} after accounting for the loss in the fibre. We observe broadening of the pump spectrum, and more importantly, appearance of a prominent mid-IR spectral band centred at \SI{3.16}{\micro\meter} in the blue-edge of the first transmission band. Its peak is approximately \SI{10}{\decibel} below the main pump spectrum and the energy in the mid-IR calculated from the intensity calibrated spectrum is \SI{1.1}{\micro\joule}. This amounts to \SI{8.6}{\percent} energy conversion from the input. Both the total and mid-IR beams at the output have near-Gaussian profiles as shown in the inset in Fig.~\ref{fig3}a, indicating that they emerge from the fundamental mode of the fibre.

A numerical simulation of the pulse propagation is carried out for the same set of parameters. Simulated output spectrum shown in Fig.~\ref{fig3}a exhibits a high-level of agreement with the measurement, reproducing most of the spectral details including the strong mid-IR radiation and other prominent peaks around \num{1.45} and \SI{0.35}{\micro\meter}. These features appear due to phase matching between soliton and dispersive waves which leads to coherent transfer of energy from the pump \cite{Akhmediev:1995:PRA}. A false colour map of the spectral evolution along the fibre length is presented in Fig.~\ref{fig3}b. The \SI{12.5}{\micro\joule}-pump, which amounts to a higher-order soliton of order \num{2.7}, initially undergoes temporal compression and spectral broadening due to interplay between anomalous dispersion and nonlinearity. Efficient frequency conversion takes place after \SI{15}{\centi\meter} of propagation in the fibre, when low-intensity spectral tails of the broadened pump eventually extend to zero phase-mismatch points. Here, these are given by the soliton-dispersive wave phase-matching condition:
\begin{equation}
  \Delta\beta_{\textrm{DW}}=\beta-\beta_{0}-\beta_{1}\left(\omega-\omega_{0}\right)-\frac{3\omega^2}{8c^2\beta}\chi^{\left(3\right)}\lvert E_{0}\rvert^{2}=0\textrm{,}\label{eq2}
\end{equation}
where $c$ is the speed of light in free space and $\beta$ is a wavevector of the propagating mode in the fibre. $\beta_{0}$ and $\beta_{1}$ are the zeroth and first Taylor series expansion coefficients of $\beta$ at the pump, $\omega_0$, which are related to phase and group velocities, respectively. The last term in Eq.~(\ref{eq2}) is a nonlinear correction term, where $\chi^{\left(3\right)}$ is the third-order susceptibility of the filling material and $E_0$ is the peak electric field strength of the pump. Figure \ref{fig3}c is the soliton-dispersive wave phase-matching diagram that results from Eq.~(\ref{eq2}). There are two distinctive origins of phase matchings in this diagram. The first is from the general dispersion landscape of the hollow-core fibre in the central parts of the transmission bands. It is responsible for the emission at \SI{0.35}{\micro\meter}, and the equivalent capillary model also features the same phase matching. The other mechanism is caused by presence of the resonances. The rapid GVD change around each resonance creates a phase-matching point in the long-wavelength side of the resonance, i.e., in the blue-edge of the transmission band of the same order. All phase-matching points marked in Fig.~\ref{fig3}c, except for \SI{0.35}{\micro\meter}, are due to the band-edge effect. This includes the one at \SI{3.1}{\micro\meter} that is promoting the mid-IR radiation. The band-edge-induced phase matchings do not show up in the capillary model. 

\begin{figure}[h]
  \centering
  \includegraphics[scale=0.66]{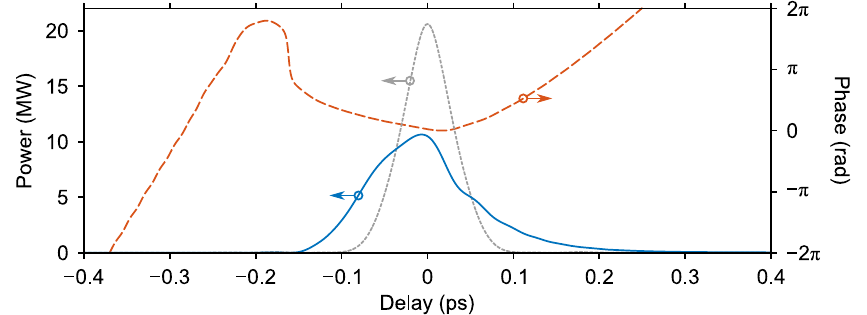}
  \caption{\textbf{Mid-IR pulse characterisation.} Temporal intensity (blue-solid line) and phase (red-dashed line) profiles of the simulated mid-IR pulse at the fibre output. The pulse is centred at \SI{3.1}{\micro\meter} wavelength, and it has FWHM duration and peak power of \SI{123}{\femto\second} and \SI{10.6}{\mega\watt}, respectively. At its Fourier limit, they are \SI{67}{\femto\second} and \SI{20.4}{\mega\watt} (grey-dotted line).}\label{fig4}
\end{figure}

We can characterise the mid-IR pulse from the numerical data. Its temporal intensity and phase profiles are shown in Fig.~\ref{fig4}. The pulse is centred at \SI{3.1}{\micro\meter} wavelength and has FWHM duration of \SI{123}{\femto\second} and peak power of \SI{10.6}{\mega\watt} at the fibre output. After its generation at around the \SI{15}{\centi\meter} mark, the mid-IR light travels linearly in the normal dispersion regime in the remaining length of the fibre. This is accountable for the frequency chirp across the pulse, which leaves a room for post pulse compression. The duration and peak power of the mid-IR pulse at its Fourier limit are \SI{67}{\femto\second} and \SI{20.4}{\mega\watt}, respectively.

\begin{figure}[h]
  \centering
  \includegraphics[scale=0.66]{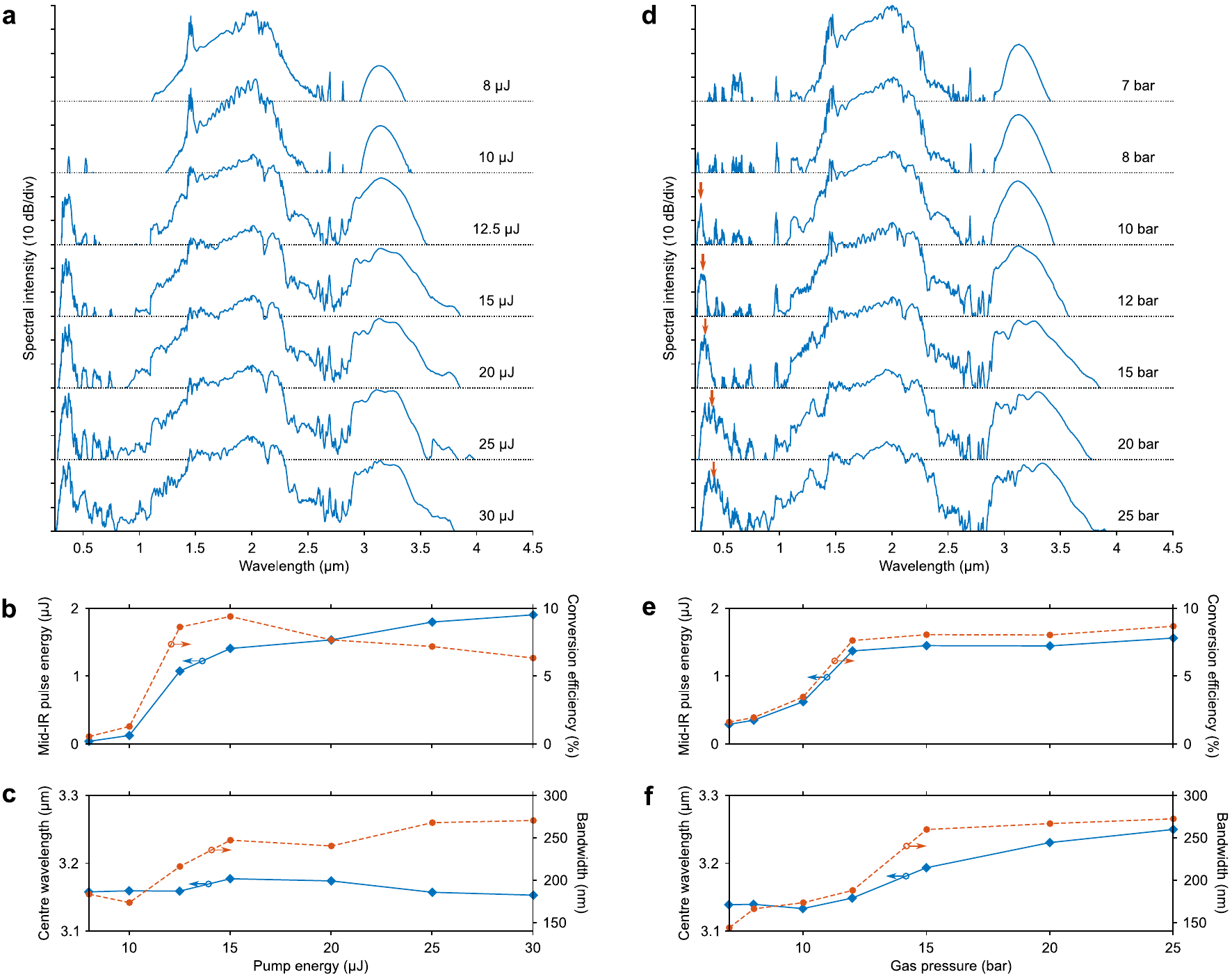}
  \caption{\textbf{Pump energy and pressure tuning.} \textbf{a.} Output spectra at different pump pulse energies when the fibre is pressurised with argon at \SI{16}{\bar}. \textbf{b.} Energy of the mid-IR pulse (blue-solid line) and the corresponding conversion efficiency (red-dashed line) versus pump energy. \textbf{c.} The centre wavelength (blue-solid line) and the FWHM spectral bandwidth (red-dashed line) of the mid-IR pulse versus pump energy. \textbf{d.} Output spectra at different argon pressurisations when the pump pulse energy is \SI{18}{\micro\joule}. The arrows mark emission due to the capillary model-induced phase matching. \textbf{e.} Energy of the mid-IR pulse (blue-solid line) and the corresponding conversion efficiency (red-dashed line) versus argon pressure. \textbf{f.} The centre wavelength (blue-solid line) and the FWHM spectral bandwidth (red-dashed line) of the mid-IR pulse versus argon pressure.}\label{fig5}
\end{figure}

One of the parameters that can be changed easily in the experiment is the pump pulse energy. It alters peak intensity of the incoming pulse which in turn affects nonlinear phase accumulation. Figure \ref{fig5}a is a collection of output spectra measured at different pump pulse energies while fixing the other settings the same as in Fig.~\ref{fig3}. The band-edge-induced spectral features at \num{3.16} and \SI{1.45}{\micro\meter} appear at the pump energy as low as \SI{8}{\micro\joule}. The latter exhibits a higher spectral intensity due to its location in the same transmission window with the pump. As the energy further increases, the main spectrum around \SI{2}{\micro\meter} broadens, and we start to observe rise of the \SI{0.35}{\micro\meter} peak brought by the capillary model-mediated phase matching. At the same time, energy in the mid-IR gradually increases. Figure \ref{fig5}b shows energy in the mid-IR spectral band, i.e., in the region above \SI{2.8}{\micro\meter}, obtained from the intensity calibrated output spectrum and the corresponding conversion efficiency versus the pump energy. The maximum mid-IR energy observed in the experiment is close to \SI{2}{\micro\joule} when the pump is \SI{30}{\micro\joule}. A monotonic increase of the mid-IR energy right up to the maximum pump used in the experiment suggests that mid-IR pulses of higher energies are achievable by further increasing the pump. The conversion efficiency, on the other hand, tops off at \SI{15}{\micro\joule}-pump where \SI{9.4}{\percent} of the input energy is converted to mid-IR. At higher pump energies, the mid-IR generation onsets at a shorter propagation distance, and hence it travels a longer length in the fibre. Since the transmission loss is higher in the mid-IR region than around the pump, the conversion becomes less efficient. The pump energy has no meaningful influence on spectral position of the mid-IR band as shown in Fig.~\ref{fig5}c. This is because the dispersion in the band-edge is largely dominated by the nearby resonance, which depends solely on the geometrical parameter, $t$. Added to that, the nonlinear correction term in Eq.~(\ref{eq2}) makes only a small contribution to the overall phase matching. Therefore, the pump energy has a negligible effect on locations of the band-edge-induced phase-matching points. On the contrary, it noticeably affects bandwidth of the mid-IR pulse. The band generally becomes broader with the rising pump energy. The enhanced broadening of the main spectrum at higher pump pulse energies leads to the conversion over a wider spectral range in mid-IR.

The gas pressure is another experimental parameter that can be varied onsite. Changing the pressure alters not only the nonlinearity but also the dispersion of the system. Figure \ref{fig5}d presents output spectra at different argon filling pressures when the pump pulse energy is \SI{18}{\micro\joule}. We observe general broadening of the main pump spectrum as well as generated spectral bands when the pressure is raised. The observation is similar to increasing the pump energy in Fig.~\ref{fig5}a, since they both enhance the system nonlinearity. The energy in the mid-IR band is plotted against the pressure in Fig.~\ref{fig5}e together with the conversion efficiency. The energy transfer to mid-IR increases rapidly when the pressure is varied from \num{7} to \SI{12}{\bar} and flattens afterwards arriving at \SI{1.6}{\micro\joule} when the pressure is \SI{25}{\bar}. One important remark is that spectral location of the mid-IR emission shifts slightly---over a bandwidth of \SI{3.2}{\tera\hertz} between \num{3.14} and \SI{3.25}{\micro\meter}---with the pressure change as evident in Fig.~\ref{fig5}f. While the resonance is the main influence on the phase matching in the band-edge, the pressure change can also cause small but observable shift in the vicinity. In comparison, emission at the phase matching supported by the capillary model, i.e., the spectral feature marked with arrows in Fig.~\ref{fig5}d, drifts across a much wider range---over a bandwidth of \SI{236}{\tera\hertz} from \num{0.31} to \SI{0.41}{\micro\meter}---when the pressure is varied from \num{10} to \SI{25}{\bar}.

\begin{figure}[h]
  \centering
  \includegraphics[scale=0.66]{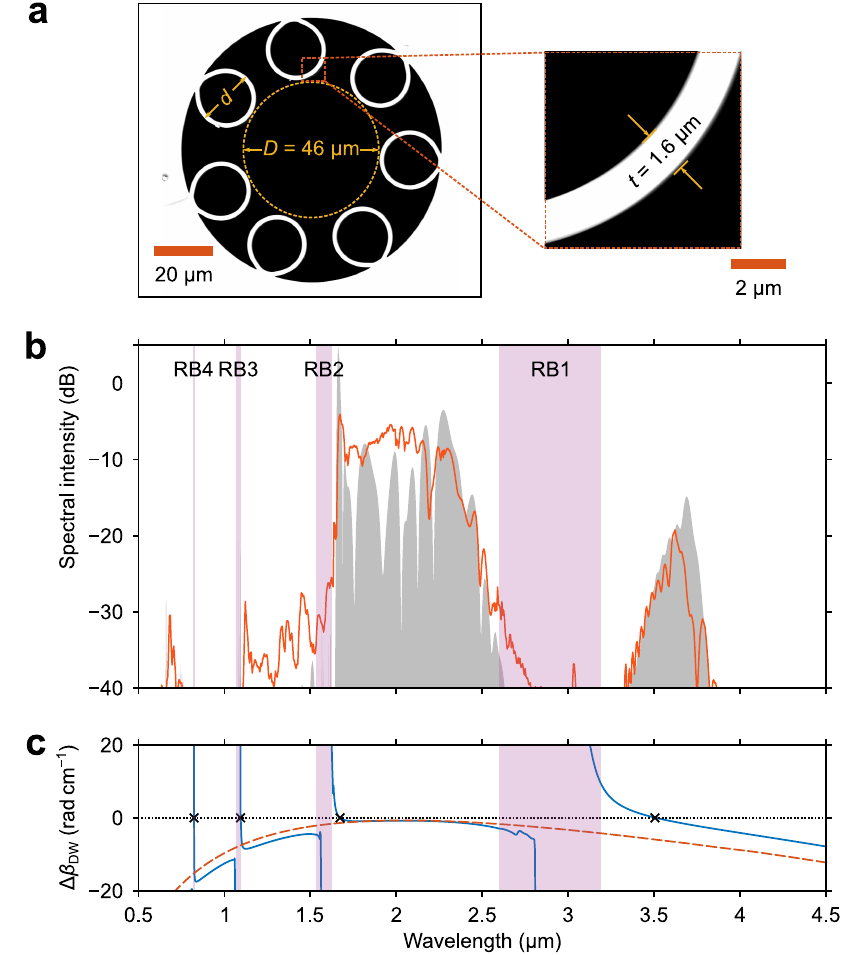}
  \caption{\textbf{Geometrical tuning of mid-IR wavelength.} \textbf{a.} A cross-sectional image of the antiresonant hollow-core fibre used in the second experiment. The right-hand side is the magnified image of a cladding element. The fibre has core diameter $D=\SI{46}{\micro\meter}$, average outer cladding element diameter $d=\SI{20}{\micro\meter}$, and dielectric wall thickness $t=\SI{1.6}{\micro\meter}$. \textbf{b.} Measured (red-solid line) and simulated (grey shade) spectra at the output of a \SI{25}{\centi\meter}-long fibre. The system is pressurised with \SI{15}{\bar} argon and the pump pulse energy is \SI{15}{\micro\joule}. \textbf{c.} The corresponding soliton-dispersive wave phase-matching diagram (blue-solid line). The dephasing of a simple dielectric capillary of the same core size and argon pressurisation is included for comparison (red-dashed line). The phase-matching points that originate from the resonant bands are marked in the diagram (black crosses). Only the first four resonant bands, denoted RB1, RB2, RB3, and RB4 (magenta shades), are considered in the dephasing diagram and band-edge induced phase-matching points for clarity.}\label{fig6}
\end{figure}

To shift the mid-IR wavelength beyond the small stretch that is attainable with the pressure tuning, we need to use an AR-HCF with a different cladding wall thickness $t$. To this end, a new set of experiments is carried out with another AR-HCF. Figure \ref{fig6}a is a cross-sectional image of the second fibre. The core and seven silica cladding tubes are smaller than the former, having diameters $D=\SI{46}{\micro\meter}$ and $d=\SI{20}{\micro\meter}$, respectively. The key parameter, $t$, is thicker at \SI{1.6}{\micro\meter}. It red-shifts the first resonant band to \SIrange[range-phrase=--,range-units=single]{2.6}{3.2}{\micro\meter}, while still allowing the \SI{2}{\micro\meter}-pump to propagate in the second transmission window. This choice of fibre is interesting, not just for studying the role of $t$, but also for understanding the impact, or lack thereof, of $D$ in the mid-IR generation. Measured spectrum at the output of a \SI{25}{\centi\meter}-long fibre filled with argon at \SI{15}{\bar} is plotted in Fig.~\ref{fig6}b. The output pulse energy is \SI{12.8}{\micro\joule} which amounts to \SI{15}{\micro\joule} at the input. Further measured data at other pump energies and argon pressures are enclosed in Supplementary Note 2. A mid-IR band appears on the long-wavelength side of the first resonant band centred at \SI{3.61}{\micro\meter}. The energy in the mid-IR band, i.e., $>\SI{3.2}{\micro\meter}$, is \SI{124}{\nano\joule}. Simulated output is generally in good agreement with the experiment. The mid-IR spectral feature especially is well reproduced. The conversion to mid-IR onsets at a shorter propagation distance in this fibre due to the smaller core size that leads to a higher launch pulse peak intensity and earlier accumulation of the nonlinear phase shift. Here, the mid-IR must travel a longer length in the fibre, exposing it to a substantially higher loss than the pump due to the small core size \cite{Deng:2021:Crystals}. This results in a significantly poorer conversion efficiency. The early formation of the long-wavelength band in the fibre is also evident from the broader temporal profile of the mid-IR pulse at the fibre output. Its FWHM duration and peak power characterised from the numerical data are \SI{254}{\femto\second} and \SI{1.1}{\mega\watt}, respectively. It can be compressed to \SI{110}{\femto\second} duration and \SI{2.8}{\mega\watt} peak power at its Fourier limit. The soliton-dispersive wave phase matching diagram of the system is presented in Fig.~\ref{fig6}c with the band-edge-induced zero-mismatch appearing next to each resonant band. Note that phase-matching due to the capillary characteristic of the dispersion is located at \SI{0.22}{\micro\meter}. Again, the resonance plays a crucial role in creating the zero-mismatch point deep in the mid-IR region. 

\section{Discussion}

The generation of microjoule-level mid-IR femtosecond pulses is demonstrated in gas-filled AR-HCFs. The technique exploits the band-edge effect, which dominates the dispersion landscape around structural resonances of the fibre. The wavelength of the mid-IR emission is dictated primarily by location of the resonance, hence highlighting the main advantage of the approach; we can induce radiation at a desired wavelength by controlling the wall thickness of the cladding elements alone. Other system parameters, such as the pump pulse energy and its duration, as well as the fibre length, core size and gas filling, are still relevant for inducing nonlinear effect to mediate the frequency mixing process. They dictate the extent of the conversion. This can be further exploited, together with the high power-handling capability of hollow-core fibres, to facilitate power-scalable mid-IR generation. 

In the experiments, the pump is deliberately chosen at \SI{2}{\micro\meter} considering the rapid advances in high-performance ultrafast fibre lasers at this wavelength. The system demonstrated herein can be integrated at the output of such a laser to realise a high-power mid-IR fibre laser that has a small footprint, near-diffraction-limited output beam quality, and exceptional stability. One interesting prospect of the source is driving ultrabroadband mid-IR comb generation. The phase-matched conversion maintains coherence between the spectral components in the pump and mid-IR. Therefore, with the right choice of the pump laser, the comb lines remain intact in the mid-IR pulses.

\section{Methods}

\subsection{Fibre characterisation}

The AR-HCFs used in the experiments are fabricated in Nanyang Technological University. The normalised transmission spectrum is obtained with a supercontinuum laser spanning \SIrange[range-phrase=--,range-units=single]{0.4}{2.4}{\micro\meter}. The broadband input is coupled into a \SI{2}{\meter}-long fibre, and the output spectrum is acquired with two optical spectrum analysers to capture the full spectral range. This is then divided by the source spectrum for the normalisation. The transmission loss of less than \SI{0.35}{\decibel\per\meter} is measured at \SI{2}{\micro\meter} with a power meter by cutting-back the fibre from \num{1.5} to \SI{0.2}{\meter}.

Guiding properties of the fundamental core modes in the AR-HCFs are calculated based on their idealised structures. They are constructed on the measured geometrical dimensions in their cross-sectional images taken with a scanning-electron microscope. From these, the complex mode indices are computed using the finite-element method over a wide spectral range. The transmission loss includes the confinement loss, absorption in the silica, and surface scattering \cite{Fokoua:2012:OE}. The linear dispersion takes into account the waveguide portion obtained from the finite-element modelling, as well as the pressure dependent argon gas contribution \cite{Boerzsoenyi:2008:AO}.

\subsection{Experiment}

A piece of AR-HCF has two gas cells attached at the ends to provide the argon pressurisation. Each gas cell has a \SI{3}{\milli\meter}-thick \ch{CaF2} window for the optical access, of which the input side window has an antireflection coating covering the \SIrange[range-phrase=--,range-units=single]{1.8}{2.2}{\micro\meter} range. The one in the output gas cell is uncoated. 

The input at \SI{2}{\micro\meter} is an idler from an optical parametric amplifier which is pumped by a Ti:sapphire laser at \SI{1}{\kilo\hertz} repetition rate. A long-pass filter is placed at the output of the optical parametric amplifier to filter out the signal at \SI{1.34}{\micro\meter}. The input energy is adjusted with a half-wave plate and thin film polarizer. The light is passed through an iris to enhance its spatial beam quality and coupled into the fibre using a plano-convex lens. All input optics have \SI{2}{\micro\meter}-coating to suppress reflections. The input pulse has FWHM duration of \SI{65}{\femto\second} just before entering the gas cell as measured in the frequency-resolved optical gating technique \cite{DeLong:1994:JOSAB} (See Supplementary Note 1). Typical coupling efficiencies are $>\SI{80}{\percent}$ for the fibre with \SI{73}{\micro\meter} core diameter and around \SI{70}{\percent} for the one with \SI{46}{\micro\meter} core diameter. 

The beam leaving the output gas cell is focused to diagnostics. The total power of the output beam is measured with a thermal power meter. The spatial profiles of the total and mid-IR output beams are characterised using the knife-edge method \cite{Araujo:2009:AO}. For the spectral measurements, we employ a fibre-coupled CCD-type spectrometer covering the \SIrange[range-phrase=--,range-units=single]{0.2}{1.1}{\micro\meter} range as well as two monochromators. One of the monochromators has a silica photodiode and Peltier-cooled lead selenide photodiode for the \numrange[range-phrase=--]{0.2}{1.1} and \SIrange[range-phrase=--,range-units=single]{1}{5.5}{\micro\meter} regions, respectively. The other monochromator has a Dewar-cooled mercury cadmium telluride photodiode that covers the \SIrange[range-phrase=--,range-units=single]{2}{12}{\micro\meter} range. The data recorded on the CCD-type spectrometer is corrected between \SIrange[range-phrase=--,range-units=single]{0.24}{1}{\micro\meter} based on its grating efficiency and photodiode responsivity. The monochromators are intensity calibrated in the region $>\SI{0.4}{\micro\meter}$ with high-power calibration light sources. The spectral measurements from these devices are cross-checked and merged. The signal-to-noise ratios of the spectrometer and monochromators are \num{30} and \SI{50}{\decibel}, respectively.

\subsection{Pulse propagation simulation}

The pulse evolution along the gas-filled hollow-core fibre is simulated by solving a single-mode unidirectional field propagation equation. It is given by \cite{Chang:2013:OL}:
\begin{align}
  \partial_z E\left(z,\omega\right)=&i\left(\beta-\frac{\omega}{v}+i\frac{\alpha}{2}\right)E\left(z,\omega\right)+i\frac{\omega^2}{2c^2 \beta}\mathcal{F}\{\chi^{\left(3\right)}E\left(z,\tau\right)^3\}\nonumber\\&-\frac{\omega}{2\epsilon_0c^2\beta}\mathcal{F}\{\partial_\tau N\left(z,\tau\right)\frac{I_P}{E\left(z,\tau\right)}+\frac{e^2}{m_e}\int_{-\infty}^{\tau}N\left(z,\tau^{\prime}\right)E\left(z,\tau^{\prime}\right)d\tau^{\prime}\}\textrm{,}\label{eq4}
\end{align}
where $\epsilon_0$ is the vacuum permittivity, $\tau$ is the time frame moving at a reference velocity, $v$, and $z$ is the propagation length. $E\left(z,\omega\right)$ is the evolving optical field in the frequency domain with $\omega$ denoting angular frequency. This is obtained by taking the Fourier transformation of the fast oscillating electric field, i.e., $E\left(z,\omega\right)=\mathcal{F}\{E\left(z,\tau\right)\}$. The linear dispersion of the fundamental core mode of the gas-filled hollow-core fibre is accounted for in $\beta$. The term containing $\chi^{\left(3\right)}$ is responsible for the third-order nonlinearity of the pressurised argon \cite{Lehmeier:1985:OC}. The last term describes photoionisation which has negligible effect for the cases studied in this work, but is nevertheless included for the sake of completeness. Here, $I_P$ is the first ionisation energy, and $e$ and $m_e$ are the electron charge and mass, respectively. For the calculation of local free-electron density, $N\left(z,\tau\right)$, a model formulated by Ammosov \textit{et al.}~is adopted \cite{Ammosov:1986:JETP}.

\backmatter

\section*{Acknowledgments}

This work is supported by the Ministry of Education, Singapore, under its Academic Research Fund Tier 2 (2019-T2-2-026).

\section*{Author contributions}

AD, TG, MIH, and WC designed the experiments. AD and TG built the experimental setup, performed the experiments, and processed the experimental data. MRAH fabricated the fibres. AD and WC carried out the numerical analyses. AD and WC prepared the first draft of the manuscript, and all authors contributed to its development. WC supervised this work.

\end{document}